# Simulation of the RAFT polymerization in 3D: steric restrictions and incompatibility between species

*Alexey A. Gavrilov*[*,§], *Alexander V. Chertovich*[§,#]

[§]*Faculty of Physics, Lomonosov Moscow State University, 119991 Moscow, Russia*
[#]*Semenov Federal Research Center for Chemical Physics, 119991 Moscow, Russia*

*e-mail: gavrilov@polly.phys.msu.ru*

## ABSTRACT

In this work we developed a RAFT polymerization model taking into account the main reactions of the experimental RAFT process and implemented that model in dissipative particle dynamics (DPD). With a help of a kinetic model based on the same reaction routine, we investigated the question of how to simulate realistic reactions using such models. We showed that a simultaneous M-fold increase of the initiation probability $p_i$ and an M-fold decrease of the termination probability $p_t$ does not result in significant changes in the chain length distribution. If the RAFT/initiator ratio is small, a simplified model with no termination and immediate radical formation can be used with good enough accuracy. After that we directly compared the reaction behavior within the kinetic model and DPD. We showed that steric restrictions, which were not present in the kinetic model, can introduce noticeable changes in the system behavior. Finally, we studied the influence of the incompatibility on the RAFT polymerization process on an example classical implementation of polymerization-induced self-assembly (PISA). We showed that in systems with incompatible species number of activation-deactivation cycles does not always reflect the dispersity of the resulting chain ensemble. Moreover, we demonstrated that specifically the incompatibility between the RAFT end group and other species can have a large effect on the polymerization results.

# 1. Introduction

After 50 years since the first mention of living polymerization,[1] this technology has become widely adopted in the polymer science. Majority of the work on the synthesis of new polymers in laboratories is now carried out using controlled ("living") radical polymerization, or reversible deactivation radical polymerization (RDRP), as IUPAC recommends to call it. The wide usage of RDRP is related to the increased versatility of radical polymerization compared to ionic polymerizations, including wider monomer choice and scalable, user-friendly reaction conditions.[2] Moreover, there are good prospects for introducing this technology into real production, since a controlled process, in addition to a well-controlled molecular weight distribution, also could be used for the synthesis of block copolymers and macromolecules of complex structure.[2] It is block copolymers that are currently attracting great attention, since they allow the creation of functional microstructured materials.

When combining the ideas of microstructured materials and RDRP, one usually considers polymerization in heterogeneous conditions: polymerization-induced self-assembly (PISA) and polymerization-induced phase separation (PIPS). The most widely-used technique here is reversible addition-fragmentation chain transfer (RAFT), well described in relation to PISA[3,4] and PIPS.[5–7] However, conducting a living process in a heterogeneous environment is fraught with complexities and potential problems mostly due to the unevenness of the reaction processes in time and space. This could greatly affect the "controllability" of the process and lead to undesired alterations of the resulting systems. The ability to tune molecular weight distributions[8,9] can have a significant impact on the self-assembly of polymer chains in solution or in the solid state, resulting in the emergence of new morphologies and material properties.[10] While a significant amount of research has been conducted on self-assembly of block copolymers with low dispersity, block copolymers with broader and more complex molecular weight distributions are only scarcely explored.[11] Computer simulations under these circumstances can become a great tool to predict the molecular weight distributions during heterogeneous polymerization and the influence of the resulting distributions on the system properties. Apart from the heterogeneous reactions, the detailed modeling can be helpful for investigating the case of sequence-controlled polymerization: single unit monomer insertion in RAFT has been proposed as a tool[12,13] for the precise introduction of functional units within polymer chains.

The vast majority of the models presented in the literature are based on the numerical solution of kinetic equations in the approximation of an absolutely homogeneous system, i.e. instant balancing of the concentrations of all components in space and time. The most typical example of such a model is the PREDICI software.[14,15] As an example of a more complex model, one should mention a recent work,[16] where the local concentrations of the reagents near the surface are taken into account as a transitional step from zero to three dimensions models.

Recent innovations and improvements focusing on the initiation and manipulation of the RDRP processes, like light-driven processes (so-called PET-RAFT[17,18]), have indicated that reactions can be spatially controlled by external stimuli. Moreover, in our recent work[19] we showed that the incompatibility between reacting species can significantly influence the reaction process; this problem is overlooked in the existed literature on RAFT. Particle-based computer modeling can help to give an in-depth understanding of these problems, because this is the most straightforward way to fully take into account all the described effects. However, to the best of our knowledge, there has been no studies related to the development and application of RAFT models capable of taking into account complex spatial distribution of species as well as the presence of incompatibility between them.

In this work, we developed a coarse-grained model of the RAFT polymerization. Our research is in many respects a continuation and development of our work[19], where a simplified process of living polymerization in the environment of partially compatible monomer units is considered. Now this model has been largely rethought and applied directly to RAFT, taking into account all the main features of this process. We propose an implementation of the RAFT process into dissipative particle dynamics, compare it with a kinetic model and show where it can give significantly different results. Our work is rather methodological, discussing application of the model to realistic system to open a new direction in the study RAFT-based systems using computer simulations.

## 2. Methods

In this work we studied the RAFT polymerization process using two approaches: dissipative particle dynamics with an implemented reaction procedure and a kinetic model in the sequence space. In both these approaches, a reaction scheme based on the main steps of realistic RAFT process was utilized; all the simulated reactions are presented in Fig.1. Typical radical initiators upon decomposition produce 2 radicals; in the employed model, however, every initiator produces 1 radical instead of 2. This simplification may be considered as mere averaging over time however, since the efficiency of the typical diazo-initiators is in fact usually around 0.5.[20,21] For the termination reaction, we took into account only the recombination.

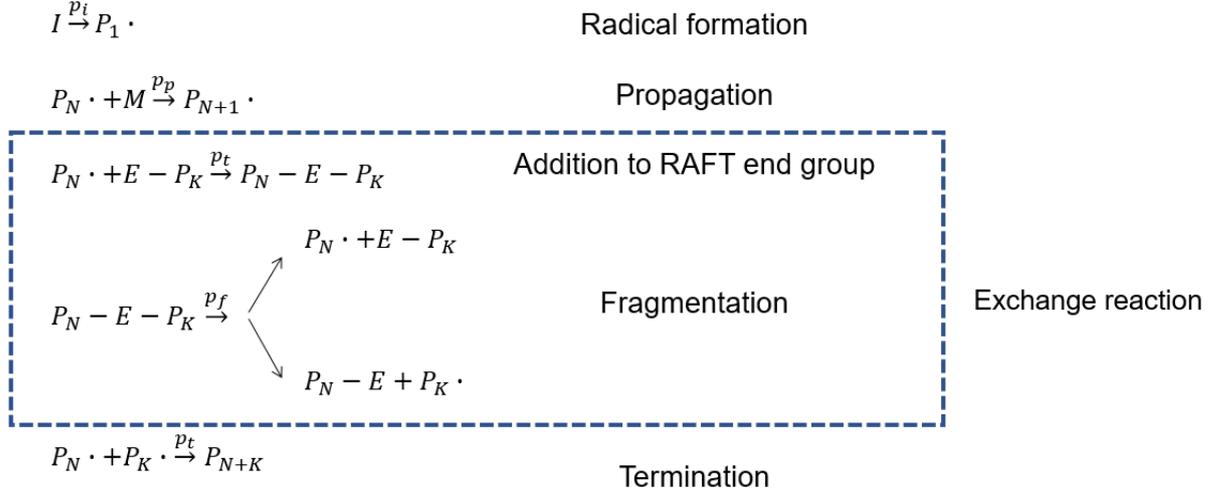

Fig.1 Scheme of the reactions included in the models.

In this scheme, $I$ is initiator, $P_X$ – chain of length X, $M$ is monomer, and dot (·) represents active radical. In our work (or, more specifically, in the DPD model, see below) we explicitly take into account that RAFT agents consist of two parts:[20] R-part, which acts as initial part of growing chains, and RAFT end group, which is transferred between chains and keeps them dormant. The latter is shown as $E$ in Fig.1. In our work the probability $p_p$ does not depend on the radical type, i.e. it is the same for initiator radical, propagating chain radical as well as R-part of RAFT agent.

Let us now describe the models in more detail.

### 2.1. Dissipative particle dynamics method without reactions

First we give a brief description of the standard dissipative particle dynamics method. Dissipative particle dynamics (DPD) is a version of the coarse-grained molecular dynamics adapted to polymers and mapped onto the classical lattice Flory–Huggins theory.[22–25] Macromolecules are represented in terms of the bead-and-spring model, with beads interacting by a conservative force (repulsion) $\boldsymbol{F}_{ij}^c$, a bond stretching force (only for connected beads) $\boldsymbol{F}_{ij}^b$, a dissipative force (friction) $\boldsymbol{F}_{ij}^d$, and a random force (heat generator) $\boldsymbol{F}_{ij}^r$. The total force is given by:

$$\boldsymbol{F}_i = \sum_{i \neq j} \left( \boldsymbol{F}_{ij}^c + \boldsymbol{F}_{ij}^b + \boldsymbol{F}_{ij}^d + \boldsymbol{F}_{ij}^r \right) \tag{1}$$

The soft core repulsion between *i*- and *j*-th beads is equal to:

$$F_{ij}^c = \begin{cases} a_{\alpha\beta}(1 - r_{ij}/R_c)r_{ij}/r_{ij}, & r_{ij} \leq R_c \\ 0, & r_{ij} > R_c \end{cases}, \quad (2)$$

where $r_{ij}$ is the vector between *i*-th and *j*-th bead, $a_{\alpha\beta}$ is the repulsion parameter if the particle *i* has the type $\alpha$ and the particle *j* has the type $\beta$ and $R_c$ is the cutoff distance. $R_c$ is basically a free parameter depending on the volume of real atoms each bead represents;[25] $R_c$ is usually taken as the length scale, i.e. $R_c=1$. In our simulations we used $a_{\alpha\alpha}=25$. In this case, the interaction parameters $a_{\alpha\beta}$ and a more common Flory-Huggins parameter $\chi$ are linearly related to each other:[25]

$a_{\alpha\beta} = \chi/0.286 + 25, \quad \alpha \neq \beta.$

If two beads (*i* and *j*) are connected by a bond, there is also a simple spring force acting on them:

$$F_{ij}^b = -K(r_{ij} - l_0)\frac{r_{ij}}{r_{ij}}, \quad (3)$$

where $K$ is the bond stiffness and $l_0$ is the equilibrium bond length. We do not give here a more detailed description and parameters discussion of the standard DPD scheme, it can be found elsewhere.[25] In our simulations we used the following set of parameters: $a_{\alpha\alpha}=25$, $K=4$, $l_0=0$.

## 2.2. Implementation of the RAFT process in DPD

The implementation of the reaction procedure in various particle simulations has been done in a number of works.[19,26–30] However, the polymerization models in the works dealing with RDRP are usually better suited to describing ATRP and NMP processes where the chain ends change their state (from dormant to active and vice-versa) virtually independently of each other. In the RAFT process, however, the RAFT end group is transferred directly between chains upon their contact.

To simulate the RAFT process in DPD, we use a widely-used Monte Carlo scheme, and the reaction routine runs after each $\tau_0=10$ DPD integration steps. This routine consists of the following stages:

1. Every previously non-initiated initiator becomes a radical with the probability $p_i$.

2. A chain end (i.e. chains of any lengths starting with initiators or R-parts of the RAFT agents) is picked at random. If the chosen end is an active radical, a list of its neighbors (located within the distance of $R_{chem}=R_c=1.0$, i.e. the potential cutoff distance, see above) is created.

3. The closest untested neighbor to the chosen is picked after that. Several situations are possible then:

   a. The chosen "neighbor" cannot react with the radical, i.e. if it is, for example, a solvent particle or a monomer unit.
   b. The chosen "neighbor" is a monomer. In this case, the propagation reaction can occur with the probability $p_p$.
   c. The chosen "neighbor" is a RAFT end group. In this case, the addition of the radical to the RAFT end group can occur with the probability $p_r$, and an intermediate radical is formed.
   d. The chosen "neighbor" is another active radical. In this case, the termination reaction can occur with the probability $p_t$.

4. If none of the reactions described in 3b-d occur, then we go back to stage 3 and pick the next closest neighbor.
5. When there are no neighbors left or a reaction described in 3b-d occur, we go back to stage 2. The stage 2 is repeated K times overall, where K is the total number of chain ends. This ensures that on average every chain end is tested once per reaction routine.
6. The fragmentation reaction takes place for each intermediate radical with the probability $p_f$. The chain which obtains the radical after the fragmentation is chosen randomly independently of its type. This reaction is assumed to be fast, and we took $p_f=0.1$ which ensured small lifetime of intermediate radicals.

For brevity, we will call the combination of reactions 3c+6 an exchange reaction. We would like to note that in our work we take into account that the fragmentation reaction can go not only in the "forward" direction, i.e. when the RAFT end group is transferred to the new chain, but also "backwards". This means that the "effective" chain exchange rate $C_{ex}$, which shows how much faster the exchange reaction is compared to the propagation reaction, in our model is calculated as $C_{ex} = \frac{0.5 \times p_r}{p_p}$.

## 2.3 Kinetic model in the sequence space

In order to investigate the RAFT polymerization in an efficient manner and be able to study a wide range of parameters and conditions, a model which produces the chain sequences according to the reaction rules, but do not take into account the spatial positions of the species in the system. Within our model, the concentrations of the species around the propagating radicals are always equal to the average in the system; this means that the system is always ideally mixed and no steric restrictions are present. The former corresponds to the situation when all the reactions are in fact kinetically controlled, which is a simplification as the termination reaction is often diffusion controlled.[31]

In general, the reaction scheme is the same as in the DPD model described above with two minor differences. First, during the stages 2 of the reaction, the neighbors list is filled with completely random particles; any particle could be chosen as a "neighbor" of the chain end, including solvent or monomer units in chains. This corresponds to the assumption that the system is ideally mixed between reactions and no steric restrictions are present. The neighbors list size was chosen to 12, since we found that in DPD beads on average have ~12 neighbors for the chosen set of simulation parameters; this way, a direct comparison between the models is easier. The second minor difference is related to the fact that the RAFT end groups (*E* in Fig.1) are simulated implicitly, and the state of a chain end (i.e. active of dormant) is transferred as a flag.

We would like to emphasize that this model takes into account that the reaction volume is finite, and the species exhaust as the copolymerization process proceeds.

## 3.Results and discussions

### 3.1 Reaction parameters studied with the kinetic model

First of all let us discuss the main problems of the simulation of radical polymerization in particle-based methods. Due to the very high reaction rate of the chain termination reactions and rather slow initiation, the concentration of active radicals in the system is very low. Indeed, let us consider the case of homopolymerization of butil acrylate with azobisisobutyronitrile (AIBN) as initiator. When the reaction equilibrium is reached, the initiation speed is equal to the termination speed. Therefore, we can write the following equation:

$$fk_d c_i = k_t c_r^2 ,$$

*f* is the initiator efficiency (we will take it equal to 0.5[20,21]), $k_d$ is the initiator decomposition rate coefficient, $c_i$ is the initiator concentration, $k_t$ is the termination rate coefficient and $c_r$ is the active radicals concentration which we want to estimate. If the take the values for $k_d$ and $k_t$ from the literature, $k_d=1.925\times10^{-5}$ s$^{-1}$ at 65°C[32] and $k_t=2.34\times10^9$ L mol$^{-1}$ s$^{-1}$ at 60°C[33], then for the initiator concentration of $c_i=10^{-2}$ mol L$^{-1}$ we get:

$$c_r = \sqrt{\frac{fk_d c_i}{k_t}} \approx 6\times10^{-9} \text{mol L}^{-1}$$

This means that the fraction of active radicals is only $c_r/c_i=6\times10^{-7}$. From the perspective of particle-based simulations this means that in order to have 1 active radical in the system during polymerization on average, such system should contain at least $\approx 1.7\times10^6$ initiators. Moreover, since having only 1 active radical would actually result in incorrect kinetics of the termination reaction (for which obviously 2 active radicals are necessary) and the number of active radicals fluctuates, its average number should be somewhat higher to avoid the situation when only individual chains are present in the system. Therefore, it's necessary to have at least >10$^7$ initiators in the system; since initiators constitute a small fraction of the species present on the system, we are looking here at the total number of molecules in the system >10$^9$. In addition to that, polymerization is often carried out in a solvent, further increasing the system size to >10$^{10}$ particles (coarse-grained!), which nowadays is an unrealistically high number for a routine simulation study for which usually many long runs are necessary. We should note that here we do not take into account the dependence of the termination speed on the chain length,[31] but even a 100-fold decrease in $k_t$ would result in only a 10-fold decrease of the necessary system size. This argument is valid for RAFT polymerization as well since the concentration of the active radicals in a RAFT process is not higher than that in an equivalent "standard" radical polymerization in the absence of RAFT agent.

Therefore, we need to develop a strategy to efficiently simulate the RAFT polymerization without the need to use huge systems. A natural benchmark for polymerization is, obviously, the resulting chain-length distribution. Such distribution for convenience can be divided into two parts: chains starting with initiators and chains staring with R-part of the RAFT-agents. Let us study what parameters can be changed in order to increase the number of active radicals in the system by a factor of M, but still obtain the same (or close to that) chain-length distribution. From the analysis of the reaction scheme, we can assume that this can be achieved by a simultaneous M-fold increase of the initiation probability $p_i$ and an M-fold decrease of the termination probability $p_t$. Indeed, this results in the M-fold increase of the number of active radicals in the system, and, therefore, the time necessary to reach some given conversion is reduced by a factor of ~M. This means that by the time this conversion is reached the same number of initiators has been activated. Moreover, the chains stay in the active state the same time as in the unmodified system since the fraction of time it is in the active state is M times larger due to the increased number of active radicals in the system, but the overall reaction time is ~M times smaller.

Let us investigate this hypothesis using the kinetic model. To that end, the following system was studied: initiator particles fraction is 0.0006, RAFT agent particles fraction is 0.003 (i.e. the RAFT/initiator ratio is equal to 5), monomer particles fraction is 0.1434, and the rest ~0.85 is solvent. The average chain length at full conversion is therefore 40. In the reference system we took $p_i = 4\times10^{-9}$, $p_p=10^{-3}$, $p_r=3.5\times10^{-2}$ (i.e. $C_{ex}=17.5$) and $p_t=0.1$. The reaction was carried out until 95% conversion was reached; the fraction of activated initiators at that conversion was found to be ~75%. The obtained chain length distributions were compared to the cases of M=40, 100 and 1000; the results are presented in Fig.2a.

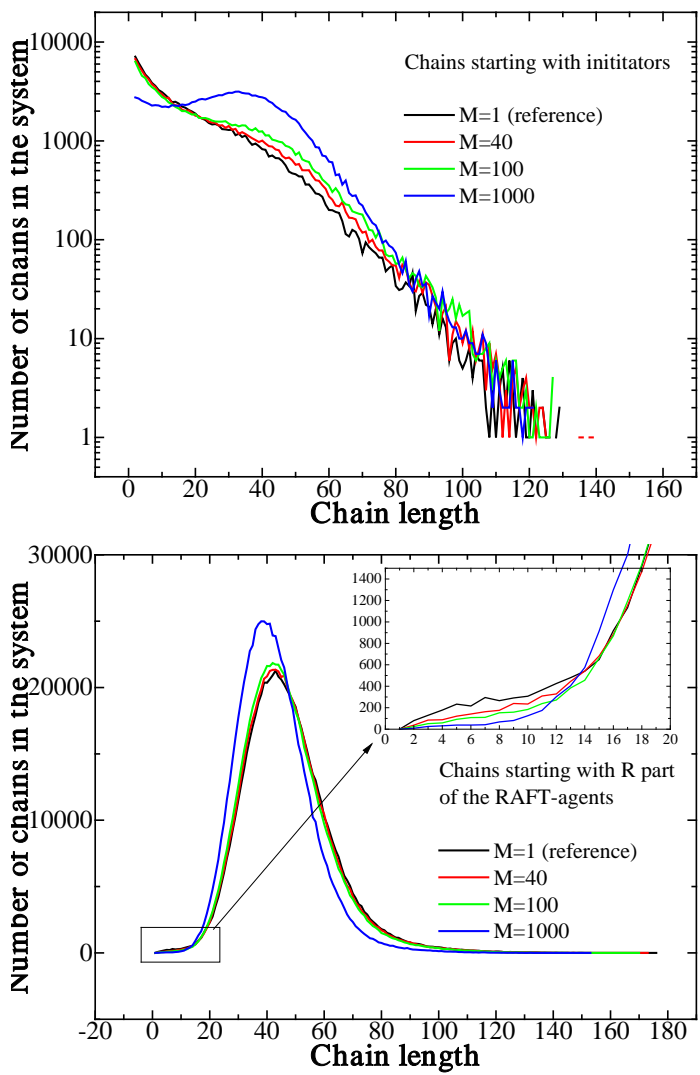

Fig.2 Chain length distributions obtained within the kinetic model for different speed increase factor M for chains: starting with initiators (top); starting with R-parts of the RAFT agents (bottom).

We see that the lengths distributions of the chains starting with initiators for M=40 and 100 have some minor differences with the reference case, while for M=1000 the differences become quite large. Similar results were obtained for the lengths distributions of the chains starting with R part of the RAFT agent (which constitute the majority of chains in the system). The large deviation from the reference case observed for M=1000 can be attributed to two factors: 1) an increase of the average length of the chains starting with initiators (see fig.2a), which decreases the average length of the chains starting with R-part of the RAFT agent; 2) smaller number of terminated chains for small values of M (due to the smaller $p_t$ as well as reaction time), which shifts the distribution peak to the left and makes it narrower. Let us further study that on a somewhat simplified case: RAFT polymerization with forbidden chain termination and immediate radical formation (i.e. $p_t=0$ and $p_i=1$), meaning that the number of active chains in the system remains constant throughout the polymerization. We studied 4 values of the fraction of active chains (which is the same is the number of initiators to the total number of chains in this case) in the system: 0.16%, 1.31%, 4.25% and 14%. For comparison, in the above-investigated "full" model the maximum fraction of active chains for M=1000 reached 11% at some point during polymerization; for M=100 this number was 2.3%, and for the reference case (M=1) – 0.03%. The resulting chain-length distributions obtained within the simplified model are shown in Fig.3.

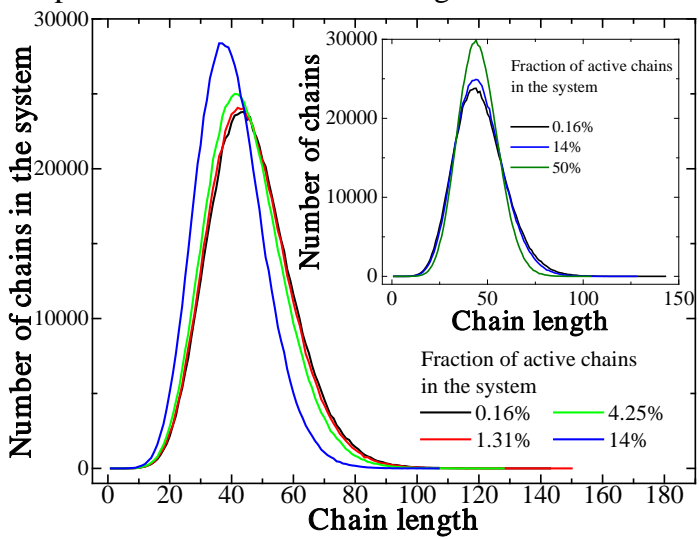

Fig.3 Chain length distributions obtained within the simplified kinetic model (no termination and immediate radical formation) for different fractions of active chains in the system. Inset shows chain length distributions obtained for systems with equal average chain length.

We see that when the fraction of active chains is small enough (0.16% and 1.31%), the distributions coincide. For larger values of that fraction, however, the distribution becomes narrower; its shift to the left is associated with the increased number of chains in the system, and, therefore, somewhat lower resulting average chain length. This can be readily demonstrated by increasing the monomer fraction in the system so that the average chain length remains constant in all cases; the resulting profiles are shown in the inset of Fig.3. We see a much better yet non-ideal correspondence between the systems with 0.16% and 14% of active chains. additionally shows an extreme case of 50% (i.e. when the number of initiators is equal to the number of RAFT agents); for that case the distribution becomes notably narrower. Such a high percentage of active chains is uncharacteristic for realistic RAFT process and presumably leads to the fact that the chain-length distribution depends less on the exchange rate $p_r/p_p$ and more on some technical details of the polymerization model which are artificial by their nature. Therefore, one should avoid such situations in simulations.

Finally, we directly compared the chain-length distributions obtained within the discussed models. We obtained that the length-distribution of the largest subset of chains – non-terminated chain starting with R-part of the RAFT agent, which are in fact the main desired product of the RAFT polymerization – is the same within the "full" (with termination and non-immediate radical formation) approach with small enough M and the simplified model with small enough constant number of active chains. This is demonstrated in the Supplemental Material, Fig.S4.

Therefore, we can draw a conclusion that the proposed approach to increasing the number of active radicals indeed can be applied without significant changes in the resulting chain length distribution; the main condition for its usage is to keep the number of simultaneously active radicals much smaller (no more than several percent) than the total number of growing chains.

Moreover, if the studied system contains only a small amount of initiator (i.e. the RAFT/initiator ratio is small), the simplified model with no termination and immediate radical formation can be used with good enough accuracy.

**3.2 Parameters of the DPD model for homogeneous systems; its comparison to the kinetic model**

As it was mentioned earlier, the kinetic model obviously does not account for possible diffusion and steric limitations present in the system as well as the influence of the incompatibility. In order to use RAFT polymerization within DPD, we must first of all define the parameters of the reaction scheme (namely, the magnitude of the reaction probabilities) which produce reliable results. The chain propagation reaction being rather slow is usually kinetically controlled.[31] We can also assume for simplicity that the same is true for the addition to the RAFT end group (unless the rate of exchange value $C_{ex}$ is extremely large). When the value of $C_{ex}>1$ (which is the most interesting case from the experimental point of view), the exchange reaction is more diffusion-dependent due to its high rate and smaller concentration of reagents compared to the propagation reaction. Therefore, we need to find the maximum $p_t$ value below which the resulting chain-length distributions do not depend on it when $C_{ex}$ is fixed. To that end, we used the simplified model with no termination and immediate radical formation for more straightforward analysis as the number of active ends does not fluctuate in that case. It is convenient to study the average total number of activation/deactivation cycles per every propagation event since in the ideal homogeneous case (see below for more discussions) is believed to control the resulting chain-length distribution.[34] It was calculated as the ratio of the total number of successful exchange reactions (i.e. those occurred in the "forward" direction) to the total number of propagation reactions at a given conversion. The following system was studied: RAFT agent fraction is 0.006 (we remind that each RAFT agent consists of 2 beads, so the total number of RAFT agents was the same as in the kinetic model studied above), monomer beads fraction is 0.1434, and the rest ~0.85 is solvent. The system size was $50^3$ (i.e. 375000 beads). We found that $p_t<=0.035$ (i.e. $p_p<=0.001$ for $C_{ex}=17.5$ for our model) seems to be a safe enough choice, at least for systems with similar concentrations of species. The details of simulations are presented in the Supplemental Material, Section A.

Let us now directly compare the behavior of the systems within the kinetic model and DPD. To that end, we compare the activation/deactivation kinetics in equivalent systems studied in the sections A and B using the simplified model for $f_a$=1.31% and $p_p$=0.001. The results are presented in Fig.4.

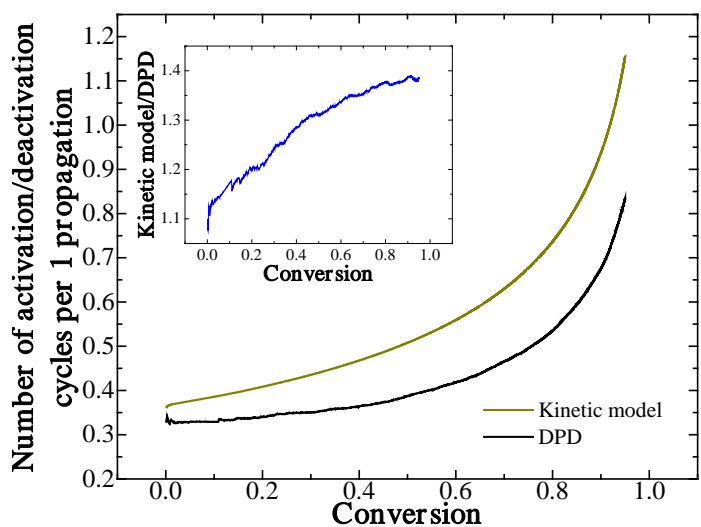

Fig.4 Comparison of the activation/deactivation kinetics obtained in simplified DPD and kinetic models for $f_a$=1.31% and $p_p$=0.001. The inset shows the kinetic model/DPD ratio.

We see an interesting feature: the number of activation/deactivation cycles is noticeably larger in the kinetic model; moreover, as the conversion increases, the difference between the kinetic model and DPD grows as it is shown in the inset of Fig.4. As it was shown above (see Fig.S1), this cannot be attributed to diffusion-related limitations due to excessively high reaction probabilities. We believe that at very small conversions the differences are explained by the fact that within the DPD model the RAFT agent consists of 2 beads, which in fact somewhat limits the access of yet point-like active chains (i.e. point-like initiators or R-parts of other RAFT-agents). Indeed, in the Methods section we mentioned that in DPD every bead has on average 12 neighbors. When a point-like active radical approaches a monomer, their access to each other is sterically unrestricted. However, for the case of the active radical–RAFT end group reaction one of the twelve neighbors of the RAFT end group is the bead it has a bond with, which reduces the total number of steric configurations available for such reaction by a factor of ~11/12. This means that within the kinetic model (in which these limitations are not present) the number of activation/deactivation cycles per one propagation should be ~12/11≈1.09 times larger than within the DPD model at very low conversions, which is what we indeed see in Fig.4. We additionally tested a modified reaction model for DPD in which the RAFT end group was imaginary, and obtained that the number of activation/deactivation cycles per one propagation is equal within such model and the kinetic model, which confirms the explanation presented above.

The subsequent growth of the difference between (inset of Fig.4) the DPD and kinetic model can be attributed to the fact that when a chain grows longer, for some chain conformations, its ends become much less accessible for an end of another chain due to the steric restrictions imposed by the monomer units of both these chains. A schematic representation of such situations is shown in Fig 5, top. It is worth noting that the observed increase in the number of activation/deactivation cycles in the kinetic model, which can basically be viewed as an increase in the "effective" $C_{ex}$ value, results only in a small change in the chain-length distribution (it becomes slightly wider). The comparison is presented in Fig.5, bottom.

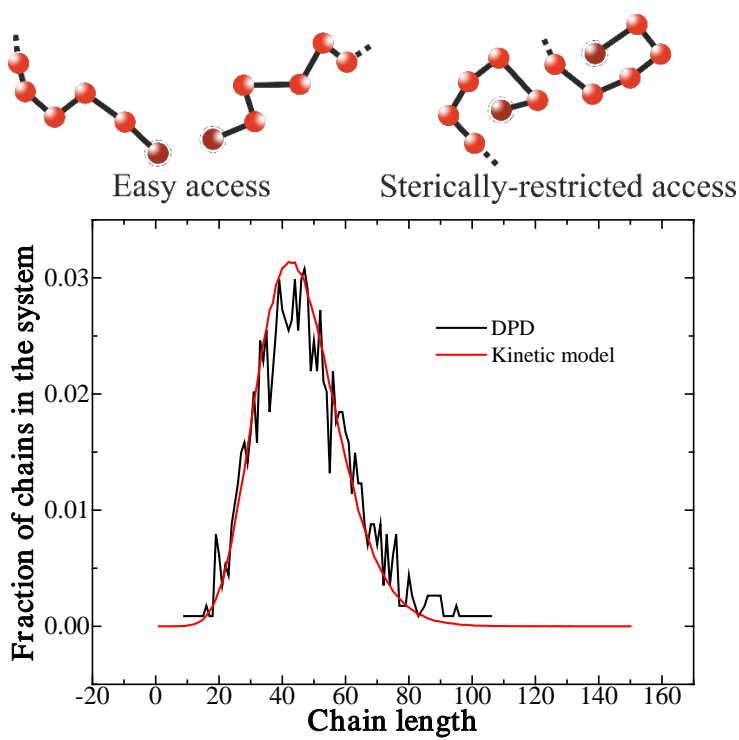

Fig.5 Schematic illustration of cases of sterically restricted and easy access of two chain ends to each other (top); Comparison of the chain length distributions obtained in simplified DPD and kinetic models for $f_a$=1.31% and $p_p$=0.001.

We should emphasize, however, that in our coarse-grained model with simple-shaped monomer units the steric restriction could be much less pronounced than for realistic polymer chains, especially with monomer units of a complex architecture, and, therefore, the resulting chain-length distribution could be influenced significantly more by the described effect.

**3.3. Heterogeneous polymerization – polymerization-induced self-assembly**

Having applied our model to homogeneous solutions, we now turn to a more realistic case of the presence of incompatibility between the species. To that end, we studied classical implementation of polymerization-induced self-assembly (PISA), for which RAFT polymerization is frequently used.[35–37] Within the standard PISA approach, second block is grown on a pre-synthesized homopolymer precursor (in our case macro-RAFT agent); the growing block is solvophobic and tends to precipitate (we will call it B-block), but the first block, which is solvophilic (A-block), stabilizes micelles.

We studied the system with the same parameters as above in the fig.4: simplified model (i.e. $p_t$=0 and $p_i$=1) with $f_a$=1.31%, monomer beads fraction 0.1434, $p_p$=0.001, $C_{ex}$=17.5. However, instead of low-molecular weight RAFT agents used in the previous section macro-RAFT agents with the length of 12 was used. The number of such macro-RAFT agents in the system was also the same as above (so the solvent content was somewhat lower), so it resulted in ≈48 monomers per one macro-RAFT agent, and the average block length ratio of the diblock-copolymers at maximum conversion would be ≈4. For simplicity, we assumed that the initiators also have type A as the soluble block.

Three cases were studied. The incompatibility between species can be characterized using Flory-Huggins χ-parameter; in the first case all the χ-parameters in the system were equal to 0. This corresponds to completely homogeneous polymerization and acted as a reference system.

Indeed, in such case the resulting chain-length distributions should be similar (neglecting the steric effects described in the previous section) to those obtained for the low-molecular weight RAFT agent. For the other two cases we introduced incompatibility between the B-type beads (monomers and monomer units) and solvent+A-type monomer units: $\chi_{B-A+S}=1.8$. Such value corresponds to high incompatibility to cause the precipitation of the growing block, but it was still not high enough to cause precipitation of the monomers at any concentration (i.e. monomer was soluble in the solvent), i.e. dispersive PISA was studied in all the cases.

The only difference between the second and third case was the behavior of RAFT-end group: in the second case it had incompatibility $\chi_{B-Reg}=1.8$ with the B-beads (and no incompatibility with the solvent and A-block), while in the third case, on the opposite, there was incompatibility between the RAFT end groups and solvent+A-type monomer units $\chi_{A+S-Reg}=1.8$ (and no incompatibility with the B-type beads). In other words, two RAFT-end groups with different side groups (and yet the same $C_{ex}$), one of which has affinity to the solvent and the other to the B-monomer, are assumed here.

First of all, let us study the activation/deactivation kinetics of these systems; the resulting dependences of the activation/deactivation cycles per every propagation event on conversion are shown in Fig.6.

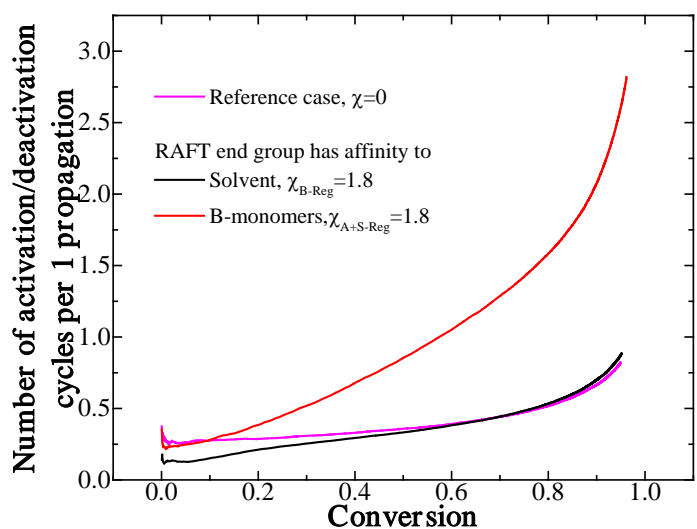

Fig.6 Activation/deactivation kinetics observed during PISA for different incompatibilities between the species in the system.

We see a large difference in the behavior: in the case when the RAFT end groups had affinity to the B-type beads, the number of exchange reactions was significantly larger than for the other two cases, and at large enough conversions this difference becomes more than 3-fold. This can be explained by the fact that the incompatibility between the RAFT end groups and the solvent leads to an increased number of contacts between the chain ends of different types, which becomes especially pronounced when the micelles start to form. On the contrary, when the RAFT end groups had affinity to the solvent, the incompatibility between them and active chain ends of the B-type resulted in somewhat reduced amount of exchange reactions compared to the reference case at low and intermediate conversions. When the micelles started to form (see fig.7) the number of contacts between the chains started to increase, so did the number of exchange reactions.

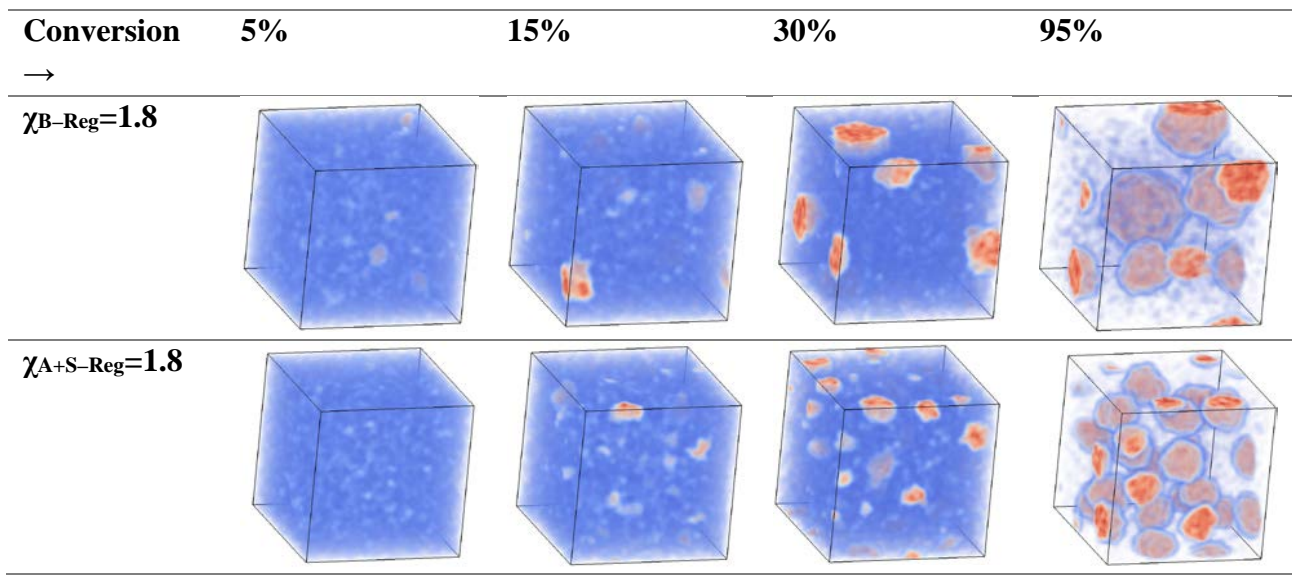

| Conversion → | 5% | 15% | 30% | 95% |
|---|---|---|---|---|
| $\chi_{B-Reg}=1.8$ | | | | |
| $\chi_{A+S-Reg}=1.8$ | | | | |

Fig.7 Snapshot of the systems (only the B-type beads are shown) at different conversions.

Let us now investigate the resulting chain-length distributions obtained at 95% conversion; they are presented in Fig.8.

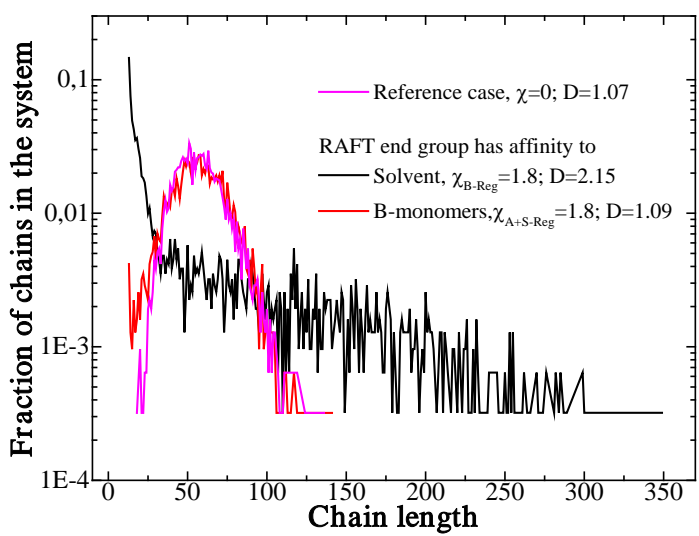

Fig.8 Chain length distributions obtained during PISA for different incompatibilities between the species in the system. Note that the chain lengths include the monodisperse soluble block A length of 12.

We see several striking features: first of all, for the case when the RAFT end groups had affinity to the B-type beads, the distribution is surprisingly somewhat wider than for the reference case despite a much larger average number of activation-deactivation cycles in the latter system (see fig.6). Next, the chain-length distribution for the case when the RAFT end groups had affinity to the solvent turned out to be extremely broad with Đ as high as 2.15, while the average number of activation-deactivation cycles was not that different from the reference case. We believe that both these results can be explained by the growing blocks affinity to each other. This in fact results in chains not freely diffusing through the system volume, but, even at low conversions when the system is still homogeneous, stay in contact with each other for some time due to the incompatibility of the growing block with the solvent. This in fact results in large number of exchange reactions which are not effective since they occur between closely located chains back and forth several times. This effect intensifies when initial stable enough aggregates start to form. Indeed, in that case the exchange reactions occur within such aggregates, resulting in the growth of the chains inside the aggregates, while the shorter chains that are still dissolved in solvent cannot grow at all since they rarely penetrate into the aggregates. This can be seen in the distributions in Fig.8: we can see that in both cases when incompatibility is present there are some unreacted macro-RAFT agents (i.e. B-clock length is 0, and the total chain length is 12), while they are not present in the reference case.

One can see an additional feature in Fig.7 related to the micelles size. Indeed, the micelles obtained in the system when the RAFT end groups had affinity to the B-type beads ($\chi_{A+S-Reg}$=1.8) are significantly smaller than those in the system with the RAFT end groups had affinity to the solvent ($\chi_{B-Reg}$=1.8), even though the average B-block length is equal for those two systems. We believe that such behavior is related to the significantly different block-length distributions (Fig. 8); indeed, in our recent work[38] we showed that even slight polydispersity in the insoluble block can noticeably change the phase behavior of diblock-copolymers in solutions. In the case of $\chi_{B-Reg}$=1.8, there is a significant amount of very long B-blocks which occupy the centers of the micellar cores, which allows them to increase their size to reduce the total interface area. The influence of the reaction conditions on structures obtained during PISA will be addressed in detail in our future works.

We additionally tested slower reaction speed, different fractions of active chains as well as the behavior within the "full" model for the case when the RAFT end groups had affinity to the solvent to see whether the results depend on the technical parameters of the model. The results were found to be consistent, indicating that they are likely to be reliable. The details are presented in the Supplemental material, section C.

To summarize, we showed that the incompatibility between the species indeed can significantly alter the resulting chain-length distributions and should be taken into account when analyzing the experimental data. Moreover, we demonstrated that the RAFT end group has a huge effect on the polymerization results not only because of the variation of $C_{ex}$ but also due to the presence of incompatibility with other species; changing its side groups to adjust its incompatibility could be an additional tool to control the reaction. Another important result is related to the fact that the number of activation-deactivation cycles does not always reflect the dispersity of the resulting chain ensemble.

**Conclusions**

In this work we developed a RAFT polymerization model taking into account the main reactions of the experimental RAFT process and implemented that model in dissipative particle dynamics (DPD). With a help of a kinetic model based on the same reaction routine, we investigated the question of how to simulate realistic reactions using such models; due to the very low concentration of active radicals in experimental systems it is problematic to do in a straightforward way. In order to increase the concentration of active radicals in the system while obtaining the same chain length distribution, we introduced the speed-up factor M. We showed that a simultaneous M-fold increase of the initiation probability $p_i$ and an M-fold decrease of the termination probability $p_t$ indeed does not result in significant changes in the chain length distribution. The main condition for the usage of this approach is to keep the number of simultaneously active radicals much smaller (no more than several percent) than the total number of growing chains. Moreover, if the studied system contains only a small amount of initiator (i.e.

the RAFT/initiator ratio is small), a simplified model with no termination and immediate radical formation can be used with good enough accuracy.

Next, we studied the question of how to choose the reaction probabilities in order to simulate kinetically controlled reactions. After that we directly compared the reaction behavior within the kinetic model and DPD. We showed that steric restrictions (which were not present in the kinetic model) can introduce noticeable changes in the system behavior. Indeed, the number of activation-deactivation cycles, which is directly connected to the dispersity of the resulting polymers, was found to be larger for the kinetic model, and the difference between the kinetic model and DPD grew as the conversion increased. We attributed that to the fact that when a chain grows longer, for some chain conformations, its ends become much less accessible for an end of another chain due to the steric restrictions imposed by the monomer units of both these chains. This could be crucial for the synthesis of polymer chains with monomer units of a complex architecture.

Finally, we studied the influence of the incompatibility on the RAFT polymerization process on an example classical implementation of polymerization-induced self-assembly (PISA). We showed that in systems with incompatible species number of activation-deactivation cycles does not always reflect the dispersity of the resulting chain ensemble. Moreover, we demonstrated that the RAFT end group has a huge effect on the polymerization results not only because of the variation of $C_{ex}$ but also due to the presence of incompatibility with other species; changing its side groups to adjust its incompatibility could be an additional tool to control the reaction. This should be taken into account when preparing and analyzing the experimental systems.

To summarize, we believe controlled radical polymerization processes still hold many intriguing questions, and computer simulations are a great tool to shed some light on them. It would be interesting to apply our model to study the RAFT polymerization in various heterogeneous media, including a detailed investigation of PISA as well as nanoreactors,[39] where all the processes are localized in a small volume and isolated from each other, and spatially controlled PET-RAFT.[17,18]

**Notes**
The authors declare no competing financial interests.

**Acknowledgment**
The research is carried out using the equipment of the shared research facilities of HPC computing resources at Lomonosov Moscow State University. The financial support of the Russian Science Foundation (project 19-73-20030) is greatly acknowledged

# Supplemental Material

**A. Comparison of the chain length distributions obtained within the "full" and simplified kinetic models.**

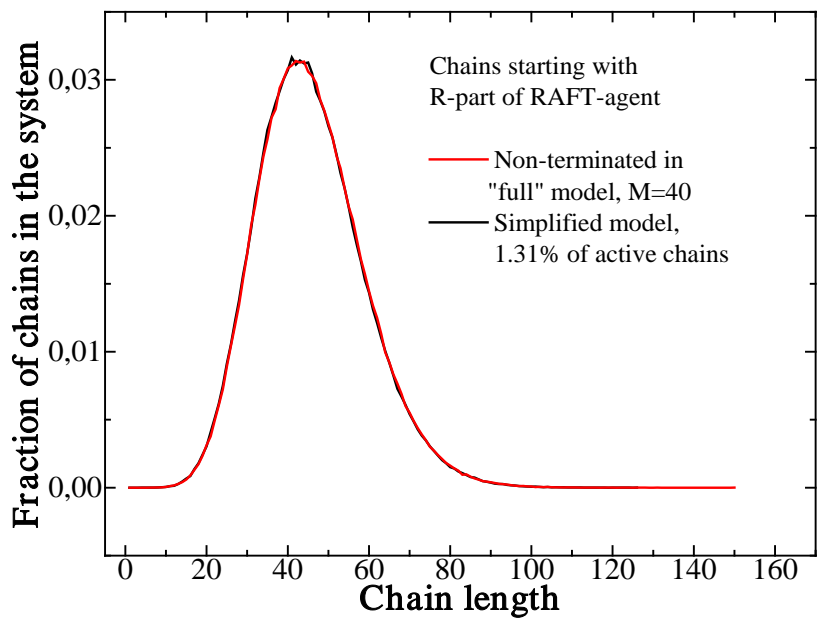

Fig.S1 Comparison of the chain length distributions obtained within the "full" and simplified kinetic models. For the full model, non-terminated chain starting with R-part of the RAFT agent are shown as the main desired product of the polymerization.

**B. Investigation of the reaction speed leading to a kinetically controlled exchange reaction in DPD**

In order to find the to find the maximum $p_t$ value below which the resulting chain-length distributions do not depend on it when $C_{ex}$ is fixed, the following system was studied: RAFT agent fraction is 0.006 (in the DPD model, each RAFT agent consists of 2 beads), monomer beads fraction is 0.1434, and the rest ~0.85 is solvent. The initiator fraction was varied; since the initiation is immediate and there is no termination, the fraction of active chains $f_a$ in the system is equal to the number of initiators divided by the total number of chains. We studied several $p_t$ values as well as the fractions of active chains $f_a$ to cover different situations, and the $C_{ex}$ value was fixed at 17.5 in all cases; the results are shown in Fig.S2.

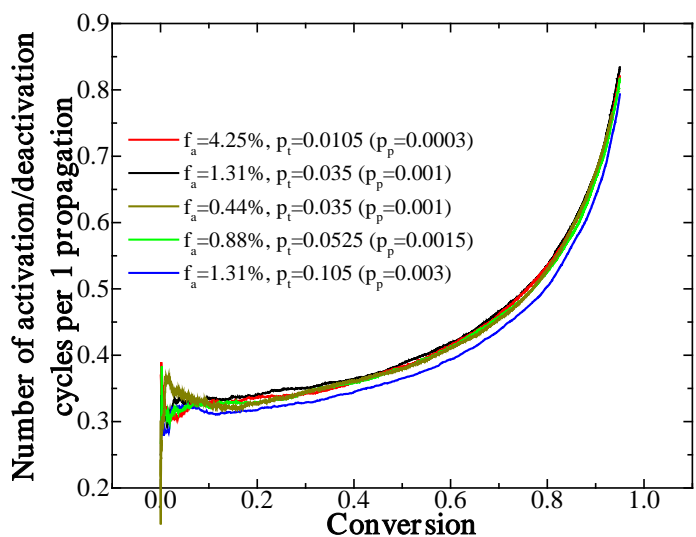

Fig.S2 Dependences of the average total number of activation/deactivation cycles per every propagation event on conversion for different fractions of active chains $f_a$ and probability of the addition to the RAFT end group reaction $p_t$ obtained using DPD.

The system with $f_a$=4.25% and $p_t$=0.0105 ($p_p$=0.0003) has the least influence of diffusion as it has the lowest reaction probabilities and, at the same time, the highest number of active chains. We see that the $p_t$=0.0525 ($p_p$<=0.0015) for all studied values of $f_a$ the curves coincide. When the value of $p_t$ becomes as high as 0.105 (corresponds to $p_p$=0.03) the number of activation/deactivation cycles starts to slightly decrease, indicating that at such probabilities the exchange reaction becomes diffusion controlled.

## C. Investigation of the reaction parameters for the case of PISA

In order to test the reliability of the results obtained for non-zero χ-values, we additionally tested slower reaction speed, different fractions of active chains as well as the behavior within the "full" model for the case when the RAFT end groups had affinity to the solvent ($\chi_{B-Reg}$=1.8, for which very wide chain-length distribution was observed). The following cases were compared:

0. Simplified model, $f_a$=1.31% and $p_p$=0.001. Identical to the system studied in the main text.
1. Simplified model, $f_a$=4.25% and $p_p$=0.0001. Compared to the system studied in the main text (case 0), this case has ~3.33 more active ends, but the reaction probabilities are 10 times smaller. This means that the reaction is significantly slower, giving the chains more time to diffuse. Also, larger number of active chains could lead to the reduction of the dispersity.
2. "Full" model with $p_i$=4×10$^{-7}$, $p_t$=0.001, $p_p$=0.001, [MacroRAFT]/[initiator]=5. In this case we found that the maximum fraction of simultaneously active chains is roughly equal to that in the case 0. The presence of the termination reaction inside the micelles could stop the growth of the longer chains forming them, and the slow radical formation from the initiators (which occurs in the solution) provides new radicals outside the micelles.
3. "Full" model with $p_i$=1.33×10$^{-7}$, $p_t$=0.003, $p_p$=0.001. In this case the simultaneous number of active chains is ~3 times smaller than in the case 2.

PISA for these 4 cases was studied in a larger simulation box of the size $100^3$ for better statistics; the obtained chain-length dependences at 50% conversion (which is well above the onset of the micelles formation, see Fig.7 of the main text) are shown in Fig. S3.

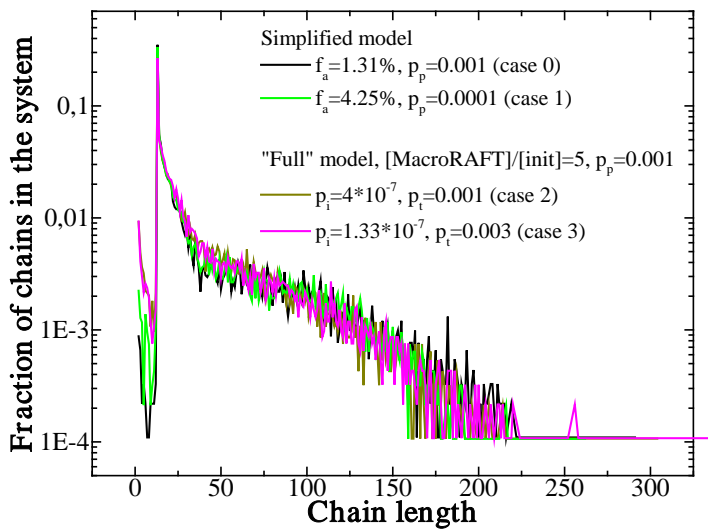

Fig. S3 Chain length distributions at 50% conversion obtained during PISA for the case when the RAFT end groups had affinity to the solvent ($\chi_{B-Reg}=1.8$) under different conditions.

We do not see any significant differences between the distributions (apart from those expected due to the apparent changes in the models).